\documentclass[review]{elsarticle}

\journal{}

\usepackage{amsmath,amssymb,amsfonts,bm}
\usepackage{url}
\usepackage{graphicx}
\usepackage{textcomp}
\usepackage{xcolor}
\usepackage{hyperref}
\usepackage{adjustbox}
\usepackage[a4paper, total={6in, 9in}]{geometry}
\usepackage{setspace}
\usepackage[croatian]{babel}

\usepackage{rotating}
\usepackage{multirow}
\usepackage{array}
\newcolumntype{R}[1]{>{\raggedleft\arraybackslash}p{#1}}
\newcolumntype{C}[1]{>{\centering\arraybackslash}p{#1}}

\date{March 31, 2019}

\begin{document}

\begin{frontmatter}

\title{Momentum and liquidity in cryptocurrencies}

\author{Stjepan Begu\v{s}i\'{c}\corref{correspondingauthor}}\ead{stjepan.begusic@fer.hr}
\author{Zvonko Kostanj\v{c}ar}
\ead{zvonko.kostanjcar@fer.hr}
\address{Laboratory for Financial and Risk Analytics \\ Faculty of Electrical Engineering and Computing, University of Zagreb \\ Unska 3, 10000 Zagreb}

\cortext[correspondingauthor]{Corresponding author \\ \indent\indent \textit{URL:} \url{lafra.fer.hr}}

\begin{abstract}
The goal of this paper is to explore the relationship between momentum effects and liquidity in cryptocurrency markets. Portfolios based on momentum-liquidity bivariate sorts are formed and rebalanced on a varying number of cryptocurrencies through time. We find a strong momentum effect in the most liquid cryptocurrencies, which supports the theories of investor herding behavior. Moreover, we propose two profitable long-only strategies: the \emph{illiquid losers} and \emph{liquid winners}, which exhibit improved risk adjusted performance over the market capitalization weighted portfolio.
\end{abstract}


\end{frontmatter}

\section{Introduction}
The cryptocurrency market has evolved rapidly throughout the past decade, attracting the interest of researchers and investment professionals seeking to understand how its diversification potential \cite{Dyhrberg2016,Platanakis2018}, underlying dynamics \cite{Ciaian2016,ElBahrawy2017}, and how their statistical characteristics \cite{Begusic2018,Phillip2018} compare to those of traditional asset classes. A study by Urquhart \cite{Urquhart2016} reports evidence that the Bitcoin market may be evolving towards efficiency, while Kristoufek \cite{Kristoufek2018} suggests that this is related to periods of cooling down after price bubbles. In addition, a study by Brauneis et al. \cite{Brauneis2018} suggests that this efficiency seems to increase with liquidity. These results are supported by Wei \cite{Wei2018} who finds lower Hurst exponents in illiquid cryptocurrencies with evidence of mean reversion. While other studies report the existence of momentum effects in cryptocurrency returns \cite{Caporale2018}, an open question remains about the pervasiveness of the momentum effect and its relation to liquidity in the cryptocurrency market. 

Motivated by results from international equity markets suggesting a stronger momentum effect in liquid market states \cite{Avramov2016} and the prospect of finding new evidence from cryptocurrency markets, in this paper we explore the relationship between momentum and liquidity using bivariate momentum-liquidity sorts over a large set of cryptocurrencies. 
We find a significant momentum effect in highly-liquid cryptocurrencies and a significant illiquidity premium within the last period's loser cryptocurrencies, supporting the line of theories which hypothesize that momentum may indeed be caused by investor herding behavior \cite{Nofsinger1999}. In addition, based on the observed phenomena, we explore two long-only cryptocurrency portfolios which exhibit notable improvement in the risk adjusted performance over the capitalization-weighted portfolio.

\section{Data}
\begin{figure}[t]
\centering
\includegraphics[width=0.75\columnwidth]{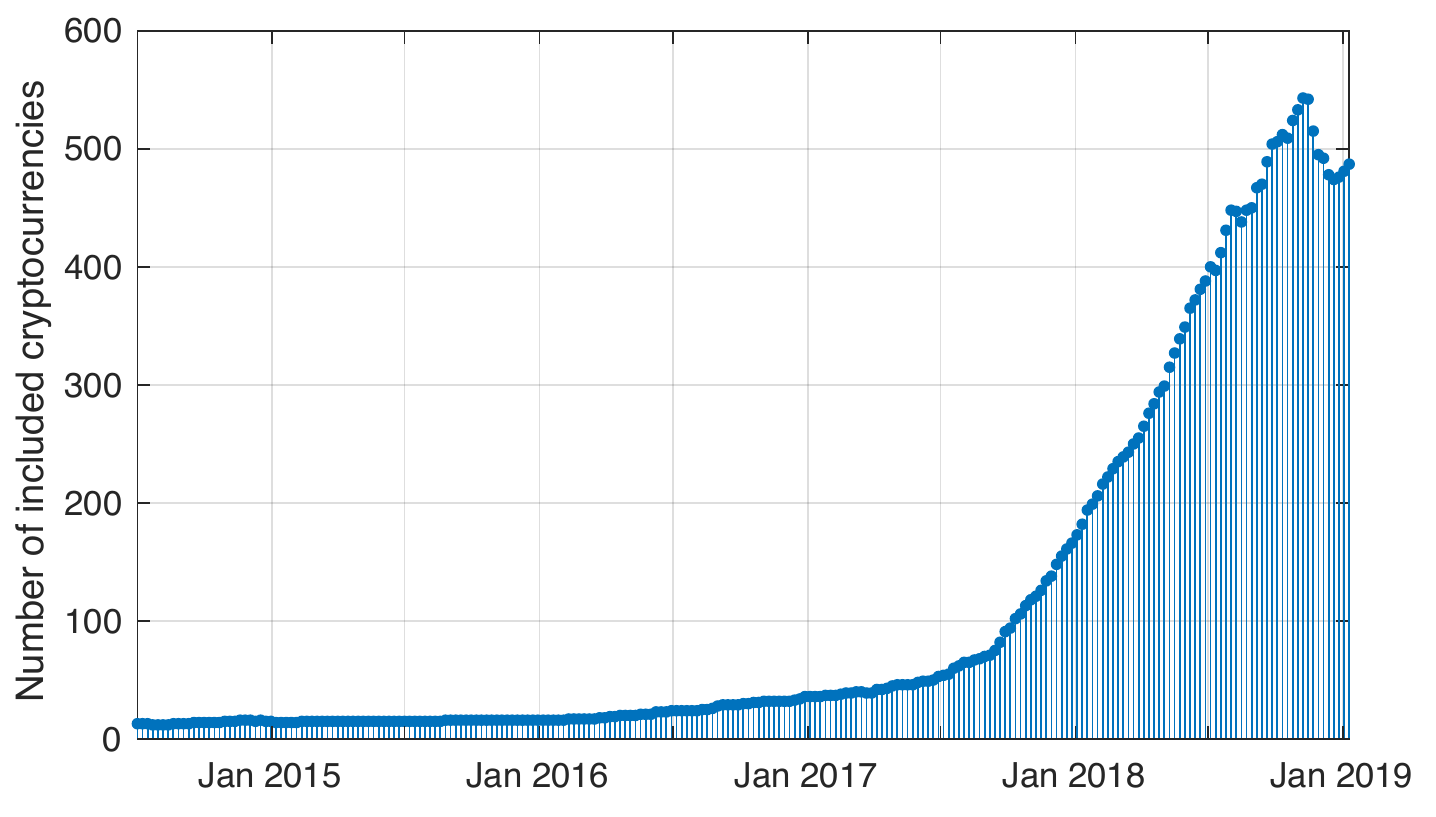}
\caption{The number of considered cryptocurrencies through time.}
\label{number_of_coins}
\end{figure}

We consider a dataset of daily cryptocurrency market capitalizations, prices, and trading volumes, obtained from \url{coinmarketcap.com}.
Firstly, we remove all of \emph{stablecoins}, which are pegged to certain fiat currencies or commodities, from the analysis. We restrict the dataset to the period from January 2015 to January 2019, since the years before 2015 contain few cryptocurrencies and very little trading activity. 
Rather than selecting a fixed subset of of cryptocurrencies, we use a set of heuristic inclusion criteria for cryptocurrencies at a given time:
(i) the cryptocurrency exists for at least a period of length $T^{(c)}$ prior to selection; (ii) the marketcap over the given period $T^{(c)}$ was never smaller than $M^{(c)}$. These conditions ensure that the most relevant and liquid cryptocurrencies are considered, just as any realistic investor would do. In this study, we use $T^{(c)} = 26$ weeks, which approximately amounts to 1/2 year, and $M^{(c)} = 10^6$ (1 million) US dollars, which is sufficiently small in order to avoid any large-cap bias. 
Throughout the analyzed time interval, a total of 711 cryptocurrencies met the proposed criteria at least once, and the number of cryptocurrencies included in the analysis through time is given in Figure \ref{number_of_coins}. 

\section{Tests and results}

%
To explore the interplay between momentum and liquidity in the cross-section of cryptocurrency returns, we form 9 momentum-illiquidity portfolios. To measure momentum, the previous performance (in terms of cummulative return) is used:
\begin{equation}
C_i(t) = \frac{P_i(t)-P_i(t-T_{\mathrm{mom}})}{P_i(t-T_{\mathrm{mom}})},
\end{equation}
over the period of $T_{\mathrm{mom}} = 14$ days (2 weeks). Based on $C_i(t)$, the selected cryptocurrencies are sorted and divided into three groups: losers (lower 30\%), neutral (between the 30\% and 70\% percentile), and winners (upper 30\%). To measure liquidity we use the Amihud \cite{Amihud2002} illiquidity measure:
\begin{equation}
I_i(t) = \frac{1}{T_{\mathrm{illiq}}}\sum_{\tau=t-T_{\mathrm{illiq}}}^t{\frac{\vert R_i(\tau)\vert}{V_i(\tau)}},
\end{equation}
where $V_i(t)$ is the traded volume (in USD) of cryptocurrency $i$ on day $t$, and the lookback period $T_{\mathrm{illiq}} = T_{\mathrm{mom}} = 14$ days. Based on this measure, the selected cryptocurrencies are divided into three illiquidity-sorted groups: liquid (lower 30\% by $I_i(t)$), neutral (between the 30\% and 70\% percentile), and illiquid (upper 30\%). The portfolios are equal-weighted and rebalanced every two weeks (14 days). We calculate the mean daily returns and their standard deviations for each of the considered portfolios. In addition, to measure the risk-adjusted performance we calculate the information ratio:
\begin{equation}
IR = \frac{\mathrm{E}\left[ R_p - R_b \right]}{\sqrt{\mathrm{var}\left[ R_p - R_b \right]}},
\end{equation}
where $R_p$ is the portfolio return and $R_b$ is the benchmark return, for which we use a capitalization-weighted portfolio,  Finally, constructed of all cryptocurrencies which satisfy the inclusion criteria at the time of rebalance, and also rebalanced every two weeks.  For each of the illiquid, neutral, and liquid classes we create zero-investment \emph{UMD = winners - losers} portfolios. Similarly, for each of the losers, neutral and winners classes we calculate the zero-investment \emph{IML = illiquid - liquid} portfolios. The statistics for momentum-liquidity bivariate sorts are given in Table \ref{momemtum_liquidity_performance}, with the results of two-tailed $t$-tests for the returns of the zero-investment $UMD$ and $IML$ portfolios.

\begin{table}[h!]
\centering

\renewcommand\arraystretch{2}
\begin{tabular}{ R{2cm} | C{2.5cm}  C{2.5cm}  C{2.5cm} | C{2.5cm} } \hline
\multicolumn{5}{c}{\textbf{Mean return [\%]}} \\ \hline
& Liquid & Neutral & Illiquid & IML  \\ \hline
Losers & 0.31 & 0.56 & 1.07 & 0.52*  \\ 
Neutral & 0.37 & 0.68 & 0.66  & 0.16 \\ 
Winners & 0.71 & 0.42 & 0.53  & -0.07 \\ \hline
UMD & 0.26* & -0.06 & -0.33  & \\ \hline \hline
\multicolumn{5}{c}{\textbf{Standard deviation [\%]}} \\ \hline
& Liquid & Neutral & Illiquid & IML  \\ \hline
Losers & 5.66 & 5.92 & 7.79  & 5.05 \\ 
Neutral & 5.18 & 6.32 & 7.36  & 4.43 \\ 
Winners & 7.43 & 6.73 & 9.45  & 7.89 \\ \hline
UMD & 5.10 & 5.25  & 8.14  & \\ \hline \hline
\multicolumn{5}{c}{\textbf{Information ratio}} \\ \hline
& Liquid & Neutral & Illiquid & IML  \\ \hline
Losers & 0.55 & 1.56 & 2.49  & 1.05 \\ 
Neutral & 0.98 & 1.92 & 1.42  & -0.09 \\ 
Winners & 1.59 & 0.82 & 0.76  & -0.55 \\ \hline
UMD & 0.22 & -0.71 & -1.08 & \\ \hline
\end{tabular}
\label{momemtum_liquidity_performance}
\caption{The mean daily return, daily standard deviation, and the annualized information ratio for the 9 momentum-liquidity portfolios, and the respective $IML$ and $UMD$ zero-investment portfolios. In the mean return table, the values related to the $UMD$ and $IML$ significant at the $\alpha = 0.05$ level according to a two-tailed t-test are marked with an asterisk (*).}
\end{table}

It is evident from the results that both the momentum ($UMD$) and illiquidity ($IML$) factors are not pervasive in all brackets. Instead, illiquidity seems to be most profitable among the losing cryptocurrencies, where the returns of the $IML$ portfolio are the highest (with a mean of 0.52\%) and statistically significant at the $\alpha=0.05$ level. Moreover, momentum seems to be most pronounced in liquid cryptocurrencies, for which the $UMD$ portfolio exhibits its highest returns with a mean of 0.26\% and is also statistically significant, while the illiquid cryptocurrencies very interestingly seem to exhibit a mean reversion effect, with the \emph{illiquid losers} portfolio exhibiting higher returns (1.07\%) than the \emph{illiquid winners} (0.53\%), which seems to confirm Wei's \cite{Wei2018} previous results. 
These findings are also confirmed by the standard deviations in the studied portfolios, which are higher in the illiquid cryptocurrencies, but also generally higher for winners. It has been argued that momentum might be caused by investor herding behavior \cite{Nofsinger1999}, and together with the positive feedback found in socio-economic signals related to cryptocurrency markets \cite{Garcia2014}, this provides an explanation for why momentum is mostly pronounced in highly-liquid cryptocurrencies.

\begin{figure}[t]
\centering
\includegraphics[width=0.75\columnwidth]{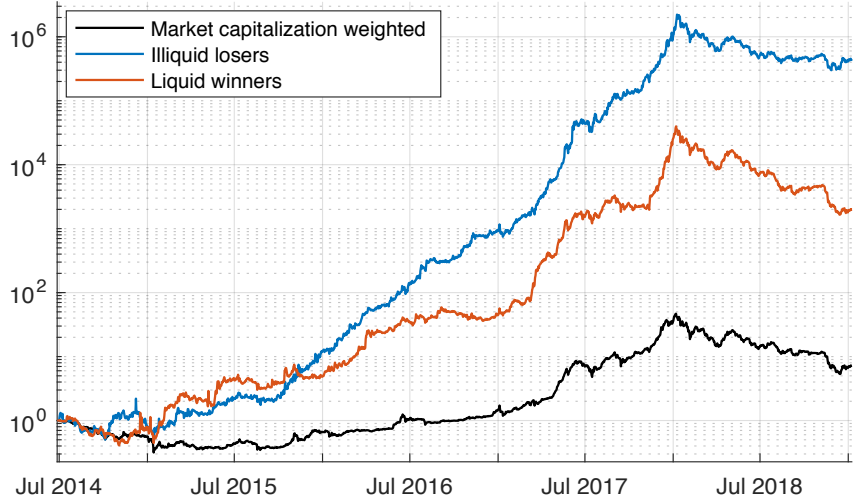}
\caption{The historical performance of the long-only \emph{illiquid losers} and \emph{liquid winners} portfolios, compared to the market capitalization weighted portfolio, all rebalanced biweekly.}
\label{portfolio_ret}
\end{figure}

Due to limited shorting opportunities in cyrptocurrency markets, the long-only \emph{liquid winners} and \emph{illiquid losers} portfolios may be the most interesting for investors. Their historic performance, without trading costs and assuming zero market impact, are shown in Figure \ref{portfolio_ret}, together with the market (capitalization weighted) portfolio. The mean return, standard deviation (volatility), and information ratios for various levels of trading costs are given in Table \ref{il_lw_portfolio_stats}.

\begin{table}[h!]
\centering

\renewcommand\arraystretch{2}
\begin{tabular}{ C{2.5cm} | C{1.5cm} C{1.5cm} C{1.5cm} | C{1.5cm} C{1.5cm} C{1.5cm}  } 
& \multicolumn{3}{c}{\emph{illiquid losers}} & \multicolumn{3}{c}{\emph{liquid winners}} \\ \hline
t. costs (bps) & $\mathrm{E}[R]$ & $\sigma$ & IR & $\mathrm{E}[R]$ & $\sigma$ & IR  \\ \hline \hline
0 &    1.07    &    7.79     &   2.49     &   0.71  &       7.43    &    1.59  \\
10 &    1.06   &     7.79   &     2.45    &     0.7  &      7.43    &    1.56 \\
50 &    1.01   &     7.79   &     2.32   &     0.65    &    7.43   &     1.43  \\
100 &    0.95    &    7.79    &    2.16    &     0.6    &    7.45     &   1.27  \\ \hline

\end{tabular}
\label{il_lw_portfolio_stats}
\caption{The mean daily return,  standard deviation (both in \%), and the information ratio for the \emph{illiquid losers} and \emph{liquid winners} portfolios, for different levels of trading costs ranging from 0 to 100 basis points (0-1\%).}
\end{table}

Although both of these long-only strategies seem profitable when met with transaction costs, the performance of the \emph{illiquid losers} portfolio may additionally deteriorate in practice, especially since it is concentrated in the most illiquid cryptocurrencies for which the indirect trading costs due to the bid-ask spread may be significantly higher. However, the \emph{liquid winners} portfolio, since it is mostly concentrated in highly-liquid cryptocurrencies, should be much less exposed to these issues in practice.

\section{Conclusion}
In this paper we explore the interaction of momentum and liquidity in the returns of 711 cryptocurrencies by forming bi-weekly rebalanced momentum-liquidity portfolios. We find a statistically significant momentum effect in the most liquid cryptocurrencies, which extends the theories of investor herding to the cryptocurrency market. In addition, we study two long-only strategies which are found to exhibit improved risk adjusted performance over the market capitalization weighted portfolio, even when transaction costs are included.

\bibliography{crypto_bk2019_bib.bib}  

\begin{thebibliography}{10}
\expandafter\ifx\csname url\endcsname\relax
  \def\url#1{\texttt{#1}}\fi
\expandafter\ifx\csname urlprefix\endcsname\relax\def\urlprefix{URL }\fi
\expandafter\ifx\csname href\endcsname\relax
  \def\href#1#2{#2} \def\path#1{#1}\fi

\bibitem{Dyhrberg2016}
A.~H. Dyhrberg,
  \href{https://www.sciencedirect.com/science/article/abs/pii/S1544612315001038}{{Bitcoin,
  gold and the dollar – A GARCH volatility analysis}}, Finance Research
  Letters 16 (2016) 85--92 (feb 2016).
\newblock \href {https://doi.org/10.1016/J.FRL.2015.10.008}
  {\path{doi:10.1016/J.FRL.2015.10.008}}.

\bibitem{Platanakis2018}
E.~Platanakis, C.~Sutcliffe, A.~Urquhart,
  \href{https://www.sciencedirect.com/science/article/pii/S016517651830274X}{{Optimal
  vs na{\"{i}}ve diversification in cryptocurrencies}}, Economics Letters 171
  (2018) 93--96 (oct 2018).
\newblock \href {https://doi.org/10.1016/J.ECONLET.2018.07.020}
  {\path{doi:10.1016/J.ECONLET.2018.07.020}}.

\bibitem{Ciaian2016}
P.~Ciaian, M.~Rajcaniova, D.~Kancs, {The economics of BitCoin price formation},
  Applied Economics 48~(19) (2016) 1799--1815 (2016).
\newblock \href {https://doi.org/10.1080/00036846.2015.1109038}
  {\path{doi:10.1080/00036846.2015.1109038}}.

\bibitem{ElBahrawy2017}
A.~ElBahrawy, L.~Alessandretti, A.~Kandler, R.~Pastor-Satorras, A.~Baronchelli,
  \href{http://rsos.royalsocietypublishing.org/content/4/11/170623}{{Evolutionary
  dynamics of the cryptocurrency market}}, Royal Society Open Science 4~(11)
  (2017) 170623 (nov 2017).
\newblock \href {https://doi.org/10.1098/RSOS.170623}
  {\path{doi:10.1098/RSOS.170623}}.

\bibitem{Begusic2018}
S.~Begu{\v{s}}i{\'{c}}, Z.~Kostanj{\v{c}}ar, H.~{Eugene Stanley}, B.~Podobnik,
  \href{https://www.sciencedirect.com/science/article/pii/S0378437118308550?via{\%}3Dihub
  http://arxiv.org/abs/1803.08405}{{Scaling properties of extreme price
  fluctuations in Bitcoin markets}}, Physica A: Statistical Mechanics and its
  Applications 510 (2018) 400--406 (jul 2018).
\newblock \href {https://doi.org/10.1016/j.physa.2018.06.131}
  {\path{doi:10.1016/j.physa.2018.06.131}}.

\bibitem{Phillip2018}
A.~Phillip, J.~Chan, S.~Peiris,
  \href{http://linkinghub.elsevier.com/retrieve/pii/S0165176517304731}{{A new
  look at Cryptocurrencies}}, Economics Letters 163 (2018) 6--9 (feb 2018).
\newblock \href {https://doi.org/10.1016/j.econlet.2017.11.020}
  {\path{doi:10.1016/j.econlet.2017.11.020}}.

\bibitem{Urquhart2016}
A.~Urquhart,
  \href{https://www.sciencedirect.com/science/article/pii/S0165176516303640}{{The
  inefficiency of Bitcoin}}, Economics Letters 148 (2016) 80--82 (nov 2016).
\newblock \href {https://doi.org/10.1016/J.ECONLET.2016.09.019}
  {\path{doi:10.1016/J.ECONLET.2016.09.019}}.

\bibitem{Kristoufek2018}
L.~Kristoufek,
  \href{http://linkinghub.elsevier.com/retrieve/pii/S0378437118302413}{{On
  Bitcoin markets (in)efficiency and its evolution}}, Physica A: Statistical
  Mechanics and its Applications 503 (2018) 257--262 (aug 2018).
\newblock \href {https://doi.org/10.1016/j.physa.2018.02.161}
  {\path{doi:10.1016/j.physa.2018.02.161}}.

\bibitem{Brauneis2018}
A.~Brauneis, R.~Mestel,
  \href{https://www.sciencedirect.com/science/article/pii/S0165176518300417}{{Price
  discovery of cryptocurrencies: Bitcoin and beyond}}, Economics Letters 165
  (2018) 58--61 (apr 2018).
\newblock \href {https://doi.org/10.1016/j.econlet.2018.02.001}
  {\path{doi:10.1016/j.econlet.2018.02.001}}.

\bibitem{Wei2018}
W.~C. Wei,
  \href{https://www.sciencedirect.com/science/article/pii/S0165176518301320}{{Liquidity
  and market efficiency in cryptocurrencies}}, Economics Letters 168 (2018)
  21--24 (jul 2018).
\newblock \href {https://doi.org/10.1016/J.ECONLET.2018.04.003}
  {\path{doi:10.1016/J.ECONLET.2018.04.003}}.

\bibitem{Caporale2018}
G.~M. Caporale, L.~Gil-Alana, A.~Plastun,
  \href{https://www.sciencedirect.com/science/article/pii/S0275531917309200}{{Persistence
  in the cryptocurrency market}}, Research in International Business and
  Finance 46 (2018) 141--148 (dec 2018).
\newblock \href {https://doi.org/10.1016/J.RIBAF.2018.01.002}
  {\path{doi:10.1016/J.RIBAF.2018.01.002}}.

\bibitem{Avramov2016}
D.~Avramov, S.~Cheng, A.~Hameed,
  \href{http://www.journals.cambridge.org/abstract{\_}S0022109016000764}{{Time-Varying
  Liquidity and Momentum Profits}}, Journal of Financial and Quantitative
  Analysis 51~(06) (2016) 1897--1923 (dec 2016).
\newblock \href {https://doi.org/10.1017/S0022109016000764}
  {\path{doi:10.1017/S0022109016000764}}.

\bibitem{Nofsinger1999}
J.~R. Nofsinger, R.~W. Sias,
  \href{http://doi.wiley.com/10.1111/0022-1082.00188}{{Herding and feedback
  trading by institutional and individual investors}}, Journal of Finance
  54~(6) (1999) 2263--2295 (dec 1999).
\newblock \href {https://doi.org/10.1111/0022-1082.00188}
  {\path{doi:10.1111/0022-1082.00188}}.

\bibitem{Amihud2002}
Y.~Amihud,
  \href{https://www.sciencedirect.com/science/article/pii/S1386418101000246}{{Illiquidity
  and stock returns: cross-section and time-series effects}}, Journal of
  Financial Markets 5~(1) (2002) 31--56 (jan 2002).
\newblock \href {https://doi.org/10.1016/S1386-4181(01)00024-6}
  {\path{doi:10.1016/S1386-4181(01)00024-6}}.

\bibitem{Garcia2014}
D.~Garcia, C.~J. Tessone, P.~Mavrodiev, N.~Perony,
  \href{http://arxiv.org/abs/1408.1494{\%}0Ahttp://dx.doi.org/10.1098/?rsif.2014.0623}{{The
  digital traces of bubbles: Feedback cycles between socio-economic signals in
  the Bitcoin economy}}, Journal of the Royal Society Interface 11~(99) (2014)
  16 (oct 2014).
\newblock \href {https://doi.org/10.1098/rsif.2014.0623}
  {\path{doi:10.1098/rsif.2014.0623}}.

\end{thebibliography}

\end{document}